\documentclass[3p,12pt,sort&compress]{elsarticle}

\usepackage{graphicx,amsmath,amssymb,amsfonts,latexsym,color,dcolumn,bm,subfigure,ulem}

\makeatletter
\def\ps@pprintTitle{%
 \let\@oddhead\@empty
 \let\@evenhead\@empty
 \def\@oddfoot{}%
 \let\@evenfoot\@oddfoot}
\makeatother

\usepackage{latexsym}
\usepackage{hyperref}
\usepackage{cleveref}
\usepackage{placeins}
\newcolumntype{L}{>{\centering\arraybackslash}m{6.5cm}}
\newcolumntype{K}{>{\centering\arraybackslash}m{13cm}}

\usepackage[colorinlistoftodos]{todonotes}
\usepackage{ulem}

\begin{document}

\renewcommand*{\thefootnote}{\fnsymbol{footnote}}

%\title{Additional evidence for an unexpected signal associated with the GW170817 binary neutron star inspiral}
\title{Comment on ``Testing claims of the GW170817 binary neutron star inspiral affecting $\beta$-decay rates''}

\author[Purdue]{E. Fischbach\corref{cor1}}
\ead{ephraim@purdue.edu}
\author[Wabash,Purdue]{D. E. Krause}
\author[Purdue]{M. Pattermann}
\address[Purdue]{Department of Physics  and Astronomy, Purdue University, West Lafayette, IN 47907, USA}
\address[Wabash]{Physics Department, Wabash College, Crawfordsville, IN 47933, USA}

 \cortext[cor1]{Corresponding author}
 
\begin{abstract}
We show that recently reported radioactive decay data  support an earlier  observation of an unexpected signal associated with the GW170817 binary neutron star inspiral.
\end{abstract}

\begin{keyword}
Neutron stars, radioactivity, neutrinos
\end{keyword}

\maketitle

In Ref.~\cite{Fischbach 2018} experimental evidence was reported for an unexpected signal associated with the GW170817 binary neutron star inspiral \cite{Abbott 1,Abbott 2}.  As noted there, this evidence derived from a laboratory experiment in which the decay rates of the isotopes Si-32 and Cl-36 were measured quasi-simultaneously in alternating time periods in a common detector.  The unexpected signal described in Ref.~\cite{Fischbach 2018}  was a surprising correlation at the $2.5\sigma$ level between the Si-32 and Cl-36 decay rates during the 5-hour period following the inspiral, which contrasts with the absence of any significant correlation either before or after this period.  In what follows we show that, when properly analyzed, data reported recently in Ref.~\cite{Breur}  exhibit a correlation similar to what was reported in Ref.~\cite{Fischbach 2018}  during the same time interval.  Given that the data reported in Refs.~\cite{Fischbach 2018} and \cite{Breur} were acquired from two experiments that were spatially separated by several thousand kilometers, the new data of Refs.~\cite{Breur} lend additional support to the indications in Ref.~\cite{Fischbach 2018} of an unexpected  signal associated with the GW170817 binary neutron star inspiral.

To place Ref.~\cite{Breur}  in proper context, there are by now numerous published reports of time-varying nuclear decay rates.  These are manifested, for example, by annual variations of decay rates attributable to the annually varying Earth-Sun distance, in a significant number of different nuclei in experiments using different techniques. (See Ref.~\cite{Fischbach 2018} for references.)  On the other hand, there have also been numerous claims that such effects have not been seen \cite{Pomme}.  What is important in evaluating the various different claims   is recognizing that since there is at present no theory to explain such effects, there is also no framework that would allow for a direct comparison of one experiment to another especially when using different nuclei.  In particular, this implies that experiments utilizing different radionuclides may have very different intrinsic sensitivities to external perturbations, and this observation is a central issue in combining the data in  Refs.~\cite{Fischbach 2018} and \cite{Breur}.  

It follows from the previous discussion that in planning a new experiment to search for possible variations in decay rates associated with astrophysical phenomena, one should carefully choose radionuclides where time-varying decay rates have previously been observed.  In Ref.~\cite{Fischbach 2018}, the radionuclides under study were Si-32 and Cl-36, both of which had previously been shown to manifest time-varying decay rates \cite{Alburger,OKeefe}.

In contrast, the same cannot be said for all the radionuclides studied in the Ref.~\cite{Breur}:  To start with, Cs-137 has been studied in great detail in Ref.~\cite{Jenkins}  based on long-term measurements carried out at the Physikalisch-Technische-Bundesanstalt (PTB) in Germany over the 9 year period 1999--2008.  The authors find that ``\ldots the PTB measurements of the decay rate of Cs-137 show no evidence of an annual oscillation\ldots''.  The same conclusion for Cs-137 arises from Ref.~\cite{Fischbach 2012}  which analyzed 5.4 years of data from a sample of Cs-137 which was on board the recent MESSENGER mission to Mercury.  Their conclusion for the relevant parameter $\xi$ was consistent $\xi =0$, i.e. no effect for Cs-137.

Turning next to Ti-44, the only directly relevant data are those of O'Keefe, et al.  ~\cite{OKeefe}.  This reference reports evidence in their Fig.~1 of an annual oscillation in the ratio of decay rates Na-22/Ti-44.  Although we cannot disentangle the respective contributions of Na-22 and Ti-44 to the ratio, we cannot rule out that Ti-44 contributes at least part of the signal for the observed annual oscillation, in which case the Ti-44 data in Ref.~\cite{Breur}  are relevant in the present context.  

Finally we consider the data reported in Ref.~\cite{Breur}  on Co-60.  This nuclide, which has been extensively studied by Parkhomov, et al. \cite{Parkhomov},  exhibits a clear annual variation, and hence the Co-60 data are also directly relevant for the present discussion.  

Turning now to Fig.~1 and Table~1 of Ref.~\cite{Breur}, which summarized their results, we can immediately set aside the Cs-137 data for the reasons discussed above.  We are then left with a figure in which the Ti-44 and Co-60 data in the relevant inspiral shaded region in Ref.~\cite{Breur}  exhibit correlations which appear to be strikingly similar to the Si-32 and Cl-36 correlations in the corresponding inspiral shaded region in Fig.~2 of Ref.~\cite{Fischbach 2018}.

\begin{figure}[t]
\centering\includegraphics[width=.9\linewidth]{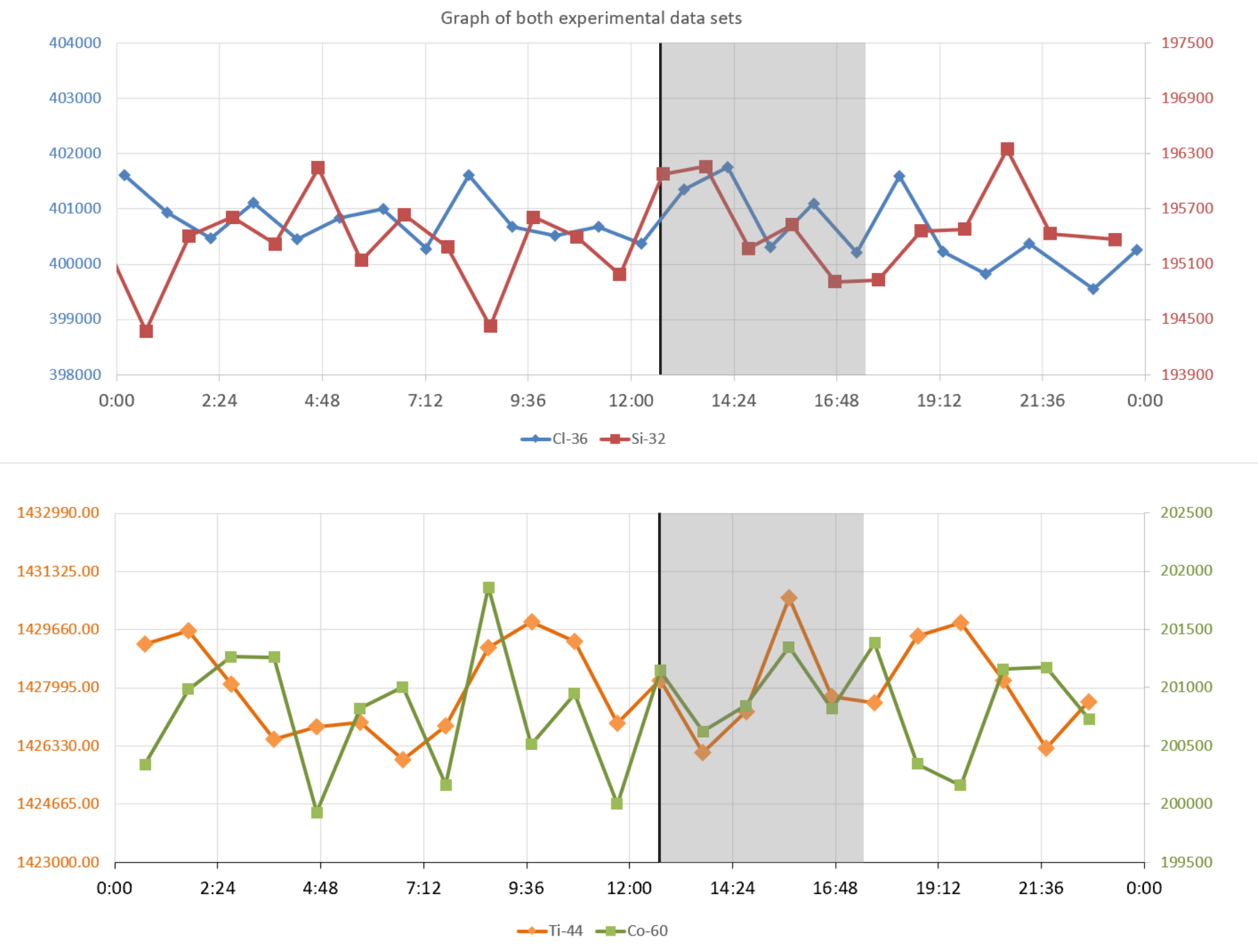}
\caption{Shown in this figure are the data from Ref.~\cite{Fischbach 2018} for Si-32 and Cl-36 (top), and the data   for Ti-44 and Co-60 obtained from the arXiv version of Ref.~\cite{Breur} (bottom).  The  gray shaded region denotes the $\sim$5-hour period following the binary inspiral, which is indicated by the vertical black line.  While there is no evidence of any significant correlations outside this 5-hour region, inside this region the Si-32 and Cl-36 data are correlated at $\sim 2.5\sigma$, while Ti-44 and Co-60 are correlated at $\sim 2.4\sigma$ as discussed in text.}
\label{graph figure}
\end{figure}

In Fig.~\ref{graph figure} below we exhibit both Fig.~2 of Ref.~\cite{Fischbach 2018}  and Fig.~1 of Ref. ~\cite{Breur}.  The data for Ti-44 and Co-60 have been obtained from the arXiv version of Ref.~\cite{Breur}, after setting aside the Cs-137 data, as discussed above. Specifically, there is now an evident correlation among the data for the four relevant isotopes (Si-32, Cl-36, Ti-44, and Co-60) which have previously been shown to evidence periodic variations in their respective decay rates.  (This correlation is already clearly evident in Ref.~\cite{Breur} itself, once the Cs-137 data are set aside.) Although the timing of the peaks and troughs in Ti-44 and Co-60 data do not correspond precisely to those in Ref.~\cite{Fischbach 2018}, this is most likely due binning effects, and to the fact that the Si-32 and Cl-36 data were acquired in the same detector but in slightly shifted time intervals.  

Although Ref.~\cite{Breur}  suggests that the authors may be disputing the effect observed in Ref.~\cite{Fischbach 2018}, the preceding analysis suggests that their data actually support the observations reported previously in Ref.~\cite{Fischbach 2018}.  Intuitively it appears unlikely that two sets of observations of a correlation in decay data in the very same $\sim$5-hour interval (but not elsewhere), obtained from disparate experiments separated by thousands of kilometers could have arisen as a result of a random statistical fluctuation.  More specifically, the comment in Sec.~4 of Ref.~\cite{Breur}  relating to a correction for the ``Look-Elsewhere'' effect is irrelevant, since the correlations that we are considering are confined to data acquired from 2 disparate experiments over exactly the  same time interval.  

In summary, it  appears that the data in Ref.~\cite{Breur}, when properly analyzed, may actually support the earlier observation published in Ref.~\cite{Fischbach 2018}  of an unexpected correlation in the decay rates of two radionuclides in the experiment during the same time interval encompassing GW170817.  If we set aside the Cs-137 data from Table~1 of Ref.~\cite{Breur}  for the reasons discussed above, the statistical significance of the Ti-44/Co-60 correlation appears to be in the range of $\sigma_{4} =(1.9$--2.4)$\sigma$, as reported in Ref.~\cite{Breur}.  Although the authors of Ref~\cite{Breur} reduce the significance of their own correlation in the post-inspiral region from an initial value of 2.4$\sigma$ to 1.9$\sigma$ because of the ``\ldots arbitrary choice of the time interval\ldots'', this reduction seems inappropriate given that this interval is, in fact, dictated by Ref.~\cite{Fischbach 2018}.  

Although the difference in the experimental setups in Refs.~\cite{Fischbach 2018}  and \cite{Breur}  preclude formally combining their respective data sets in the inspiral region to infer a combined effective significance of $\sigma_{e}$, we can presume that the joint probability of detecting these two independent sets of correlations in the same time interval has a statistical significance that is at least greater than the $\sigma_{1} =2.5\sigma$ significance quoted in Ref.~\cite{Fischbach 2018}.  (A heuristic estimate, $\sigma_{e} \sim \sqrt{\sigma_{1}^{2} + \sigma_{4}^{2}}$, would yield $\sigma_{e}\sim 3.5\sigma$ using the data of Refs.~\cite{Fischbach 2018}  and  $\sigma_{4} = 2.4\sigma.$)

We conclude by noting that since other groups may have also acquired relevant data  during the GW170817 inspiral period from nuclei that have previously exhibited decay anomalies, we encourage members of the community to revisit their data for possible signals during this period.

\section*{Acknowledgements}

We are indebted to Virgil Barnes, Laura Cayon, Tim Meese, and Jeff Scargle for helpful discussions.

\end{document}